%% file: Blois.RobertSmith.ISW.astro-ph.V1.tex
\documentclass[11pt]{article}

\usepackage{blois}
\usepackage{epsfig}
\usepackage{natbib}

\input{defs.tex}

\begin{document}


\title{Impact of scale-dependent bias and nonlinear evolution on the
  ISW}

\author{Robert E. Smith}

\address{Institute for Theoretical Physics, University of Zurich,
  Zurich, 8037, Switzerland}

\maketitle\abstracts{I summarize recent results from Smith,
  Hernandez-Monteagudo \& Seljak\,\citep{Smithetal2009}, a study of
  the impact of nonlinear evolution of gravitational potentials in the
  LCDM model on the Integrated Sachs-Wolfe (ISW) contribution to the
  cross-power spectrum of the CMB and a set of biased tracers of the
  mass. We use a large ensemble of $N$-body simulations to directly
  follow the potentials and compare the results to analytic
  perturbation theory (PT) methods.  The PT predictions match our
  results to high precision for $k<0.2 \kMpc$.  We analyze the
  CMB--density tracer cross-spectrum using simulations and
  renormalized bias PT, and find good agreement.  The usual assumption
  is that nonlinear evolution enhances the growth of structure and
  counteracts the linear ISW on small scales, leading to a change in
  sign of the CMB-LSS cross-spectrum at small scales. However, PT
  analysis suggests that this trend reverses at late times when the
  logarithmic growth rate $f=d\ln D/d\ln a<1/2$ or $\Omega_m (z)<0.3$.
  Numerical results confirm these expectations and we find nonlinear
  enhancement of the ISW signal on small scales at late-times. On
  computing the total contribution to the angular spectrum, we find
  that nonlinearity and scale dependence of the bias are unable to
  influence the signal-to-noise of the current and future
  measurements.}


\section{Introduction}\label{sec:intro}

The temperature fluctuations in the Cosmic Microwave Background (CMB)
are directly sensitive to the presence of Dark Energy, through the
change in energy that a CMB photon experiences as it propagates
through an inhomogeneous Universe with time evolving gravitational
potentials, $\pdot$. There are three main processes that give rise to
changing potentials: the Integrated Sachs--Wolfe
effect~\citep[][ISW]{SachsWolfe1967}, which deals with the linear
evolution of potentials; the Rees--Sciama
effect~\citep[][RS]{ReesSciama1968}, which is concerned with the
nonlinear evolution of potentials; and the Birkinshaw--Gull
effect~\citep{BirkinshawGull1983}, where time evolving potentials
arise due to mass flows. We define the ``nonlinear ISW effect'' as the
sum of all three contributions.

Unfortunately, the nonlinear ISW effect can not be observed directly
in the CMB auto-power spectrum, owing to the fact that it affects the
low multipoles, where cosmic variance is large. It can however be
directly observed by cross-correlating the CMB with tracers of the
Large-Scale Structure (hereafter LSS).
\citep{CrittendenTurok1996}. This analysis has recently been performed
by a number of groups using the WMAP data and several LSS measurements
(e.g. SDSS, NVSS, 2MASS), with claims of up to $4\sigma$ level
detections \citep{Giannantonioetal2008,Hoetal2008}. In a recent paper,
\citep{Granettetal2008b} measured the cross-correlation between
super-structures and super-voids with the CMB. On stacking the signal
they found a $\sim4.5\sigma$ detection, in multiple WMAP bands, and
the sign of which appeared consistent with late-time ISW. This result
is very puzzling when one considers signal-to-noise ($\SN$)
calculations within the LCDM model. These show that the maximum
$\SN\sim7$ for full sky surveys \citep{CrittendenTurok1996}. How then
is it possible to obtain such high $\SN$, given the partial sky
coverage of current surveys?  One proposed solution is that nonlinear
evolution of potentials and galaxy bias may increase the $\SN$.

In this study we calculate the impact of nonlinear evolution of
gravitational potentials on the angular cross-spectrum between a
biased set of density tracers and the CMB. We ask whether such effects
influence $\SN$. We pursue a two-pronged attack: our first line of
inquiry is analytic, and we use perturbation theory (PT) to predict
$\pdot$, and this we do in \S\ref{sec:PT}; our second line is to use a
large ensemble of $N$-body simulations to directly follow $\pdot$, and
this we do in \S\ref{sec:nbody}.


\section{The ISW effect in nonlinear PT}\label{sec:PT}

The ISW effect may be written as \citep{SachsWolfe1967}:
\be \frac{\Delta T(\nhat)}{T_0} = {2 \over c^2}\int_{t_{\rm
ls}}^{t_{0}}dt \dot{\Phi}(\nhat,\chi;t) \ \label{eq:ISW} ,\ee
where $\nhat$ is a unit direction vector on the sphere, $\Phi$ is the
dimensionless metric perturbation in the Newtonian gauge, which
reduces to the usual gravitational potential on small scales, the
`over dot' denotes a partial derivative with respect to the coordinate
time $t$ from the FLRW metric, $\chi$ is the comoving radial geodesic
distance $\chi=\int cdt/a(t)$, and so may equivalently parameterize
time. $t_0$ and $t_{\rm ls}$ denote the time at which the photons are
received and emitted (i.e. last scattering), respectively, $c$ is the
speed of light and $a(t)$ is the scale factor.

The rate of change of $\Phi$ can be calculated from Poisson's equation
($\nabla^2\Phi=4\pi G\rhob \delta a^2 $):
\be \dot{\Phi}(\bk,a) = \frac{3}{2}\Omega_{m0}
\left(\frac{H_0}{k}\right)^2\left[
  \frac{H(a)}{a}\delta(\bk,a)-\frac{\dot{\delta}(\bk,a)}{a}
  \right] \ . \label{eq:pdot} \ee 
Thus we require knowledge of the time evolution of $\delta$ and its
growth rate $\dot{\delta}$.

The collapse of cosmic structures can be followed into the nonlinear
regime using perturbation theory (PT). The solutions for $\delta$ in
Fourier space may be written \citep[][]{Bernardeauetal2002}:
\be \delta(\bk,a) = \sum_{n=1}^{\infty}[D(a)]^n\delta_n(\bk,a_0) \ \ ;  \ \ 
\delta_n(\bk) \equiv \int
\frac{\prod_{i=1}^{n}\left\{\dq_i\,\delta_1(\bq_i)\right\}}{(2\pi)^{3n-3}}
\delta^D(\bk-\bq_1-\dots-\bq_n) F^{(s)}_n(\bq_1,...,\bq_n) \label{eq:PTdelta} \ ,
\ee
where $\delta_1(\bq_i)$ represents an initial field at wavenumber
$\bq_i$, and the quantities $F^{(s)}_n(\bq_1,...,\bq_n)$ represent the
standard PT interaction kernels, symmetrized in all of their
arguments. We may obtain the time rate of change of the fluctuation
by simply differentiating the above, 
\be \dot{\delta}(\bk,a) = f(a)H(a) \sum_{n=1}^{\infty} n [D(a)]^n
\delta_n(\bk,a_0)\label{eq:PTdeltadot} \ ,\ee
where $f(a)\equiv d\log D(a)/d\log a$. Inserting \Eqns{eq:PTdelta}{eq:PTdeltadot}
into \Eqn{eq:pdot}, we find
\be \dot{\Phi}(\bk,a) = \frac{3}{2}\Omega_{m0}
\left(\frac{H_0}{k}\right)^2 \frac{H(a)}{a}\sum_{n=1}^{\infty}\left[1-
nf(a) \right]
  [D(a)]^n
\delta_n(\bk,a_0)  \ . \label{eq:pdotPT} \ee 
Consider linear theory, $n=1$: for the case $\Omega=1$, $f(a)=1$ and
potentials do not change with time and there is no ISW effect.
However, for LCDM $f(a)\approx\Omega_m^{0.6}$, and so
$1-f(a)\ge0$. Thus $\dot{\Phi}(\bk,a)>0$ for an overdensity, and
potentials decay at late times in linear theory. Consider now the
nonlinear theory, $n>1$: we immediately see that there are critical
times in the LCDM model and that these are dictated by the sign of the
bracket $\left[1-nf(a)\right]$. For $n=2$ we have $1-2f(a)<0$ for
$a<a_{\rm RS}$, and $1-2f(a)>0$ for $a>a_{\rm RS}$, and we call $a_{\rm
  RS}$ the Rees-Sciama time.  For $a<a_{\rm RS}$ nonlinear evolution
enhances the growth of potential wells, but for $a>a_{\rm RS}$ it
augments their decay. Thus it is theoretically possible to boost the
late-time ISW effect.


\section{The ISW effect in $N$-body simulations}\label{sec:nbody}

We use the Z\"urich Horizon, ``{\tt zHORIZON}'', simulations to study
the ISW. These are a large ensemble of pure CDM $N$-body simulations
($N_{\rm sim}=30$). In this study we use the first 8 simulations,
since these have 11 snapshots logarithmically spaced in the expansion
factor from $z=1$ to $0$, giving good time sampling.  Each numerical
simulation was performed using the publicly available {\tt Gadget-2}
code \citep{Springel2005}, and followed the nonlinear evolution under
gravity of $N=750^3$ equal mass particles in a cube of length
$L=1500\Mpc$. The cosmological parameters are:
$\{\Omega_{m0}=0.25,\,\Omega_{\Lambda}=0.75,
\,\sigma_8=0.8,\,n_s=1.0,\,h=0.72\}$, where these are: density
parameters in matter and vacuum energy; power spectrum normalization
and spectral index; Hubble parameter. Dark matter halo catalogs were
generated using the FoF algorithm, and the minimum number of particles
for which an object was considered bound was 30. This gave haloes with
$M> 1.5\times10^{13} M_{\odot}/h$ \citep[for more details
see][]{Smith2009}.

\begin{figure}[t]
\centering{
\includegraphics[width=6cm,clip=]{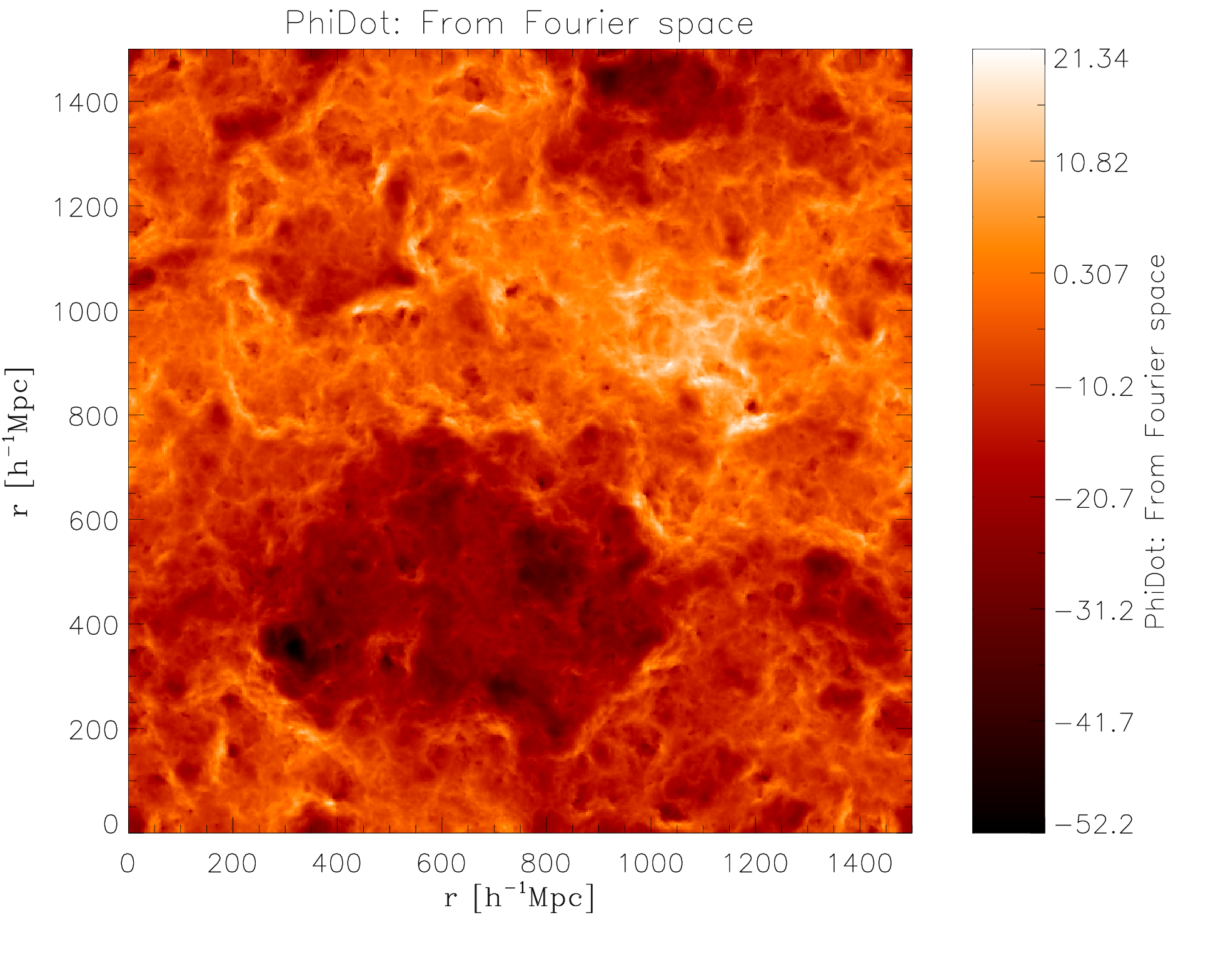}\hspace{1cm}
\includegraphics[width=6cm,clip=]{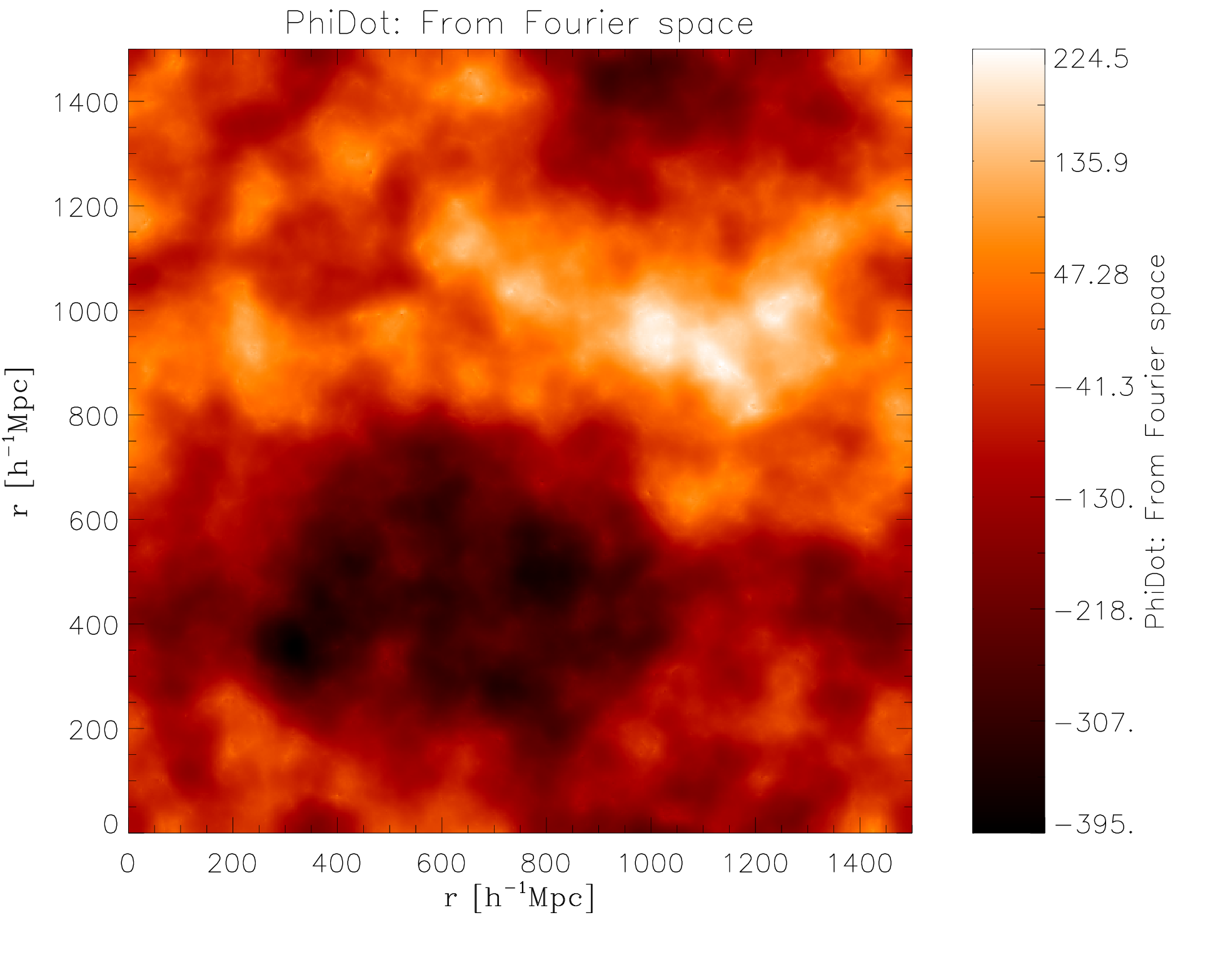}
}
\caption{\small{Evolution of $\dot{\Phi}$ in a slab of thickness
    $\Delta x=100\Mpc$. Left $z=10$ and right $z=0$.\label{fig:PhiDotMAP}}}
\end{figure}

Considering again \Eqn{eq:pdot} we see that in order to measure the
ISW in simulations we need to be able to estimate $\delta$ and
$\dot{\delta}$. The first is straightforward. The second can be
obtained from the continuity equation:
$\dot{\delta}(\bx)=-{\bf\del}\cdot\left[1+\delta(\bx)\right]{\bf
  v}(\bx)/a=-{\bf\del}\cdot{\bf p}(\bx)/a$. In Fourier space this
becomes, $\dot{\delta}_k=i\bk\cdot{\bf p}_k/a$, which may easily be
computed \citep[see][for details]{Smithetal2009}.  Figure
\ref{fig:PhiDotMAP} presents a visual representation of the ISW effect
in the simulations.


\section{Results: CMB-LSS angular cross-power spectrum}


\begin{figure}
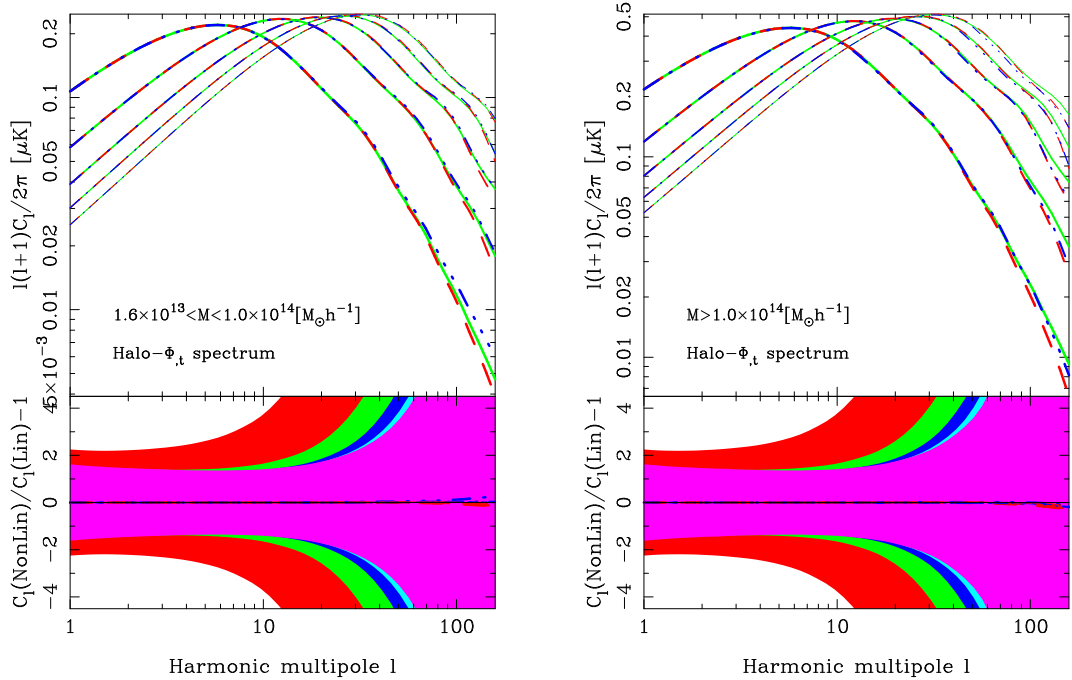

\centering{
  \includegraphics[width=6.5cm,clip=]{ClHPhi_Sim_HaloBin_2.ps_NEW}\hspace{1cm}
  \includegraphics[width=6.5cm,clip=]{ClHPhi_Sim_HaloBin_3.ps_NEW}}
\caption{\small{Angular cross-power spectrum of ISW effect and haloes
    as a function of spherical harmonic multipole $l$. {\em Left
      panel}: results for group-scale dark matter haloes. {\em Right
      panel}: results for the cluster-scale haloes. In each panel we
    show results for 5 equally spaced bins in redshift over the range:
    $z=[0.0,1.0]$. The predictions are differentiated by line
    thickness: thick lines -- low redshift; thin lines -- high. The
    line styles denote: linear theory -- solid green line; nonlinear
    PT -- red dash line; bi-cubic spline fit to the simulation data --
    blue triple-dot dash line. Top sections of each panel give the
    absolute power, and the lower sections the ratio with respect to
    linear theory. The shaded regions represents the 1-$\sigma$ error
    domains per multipole of the linear cross-spectra, where the
    central redshifts $z\in\{0.1, 0.3, 0.5, 0.7, 0.9\}$, correspond to
    the colors (red, green, blue, cyan, magenta).
\label{fig:ClHT}}}
\end{figure}


Under the Limber approximation the CMB-LSS angular cross-power
spectrum ($C_l^{\h T}$) is \citep{Smithetal2009}:
\be C_{l}^{\h T} \approx \int_{0}^{\chi_{\rm max}} d\chi\,
\frac{2a}{c^3}\Pi_{ij}(\chi)
P_{\h\pdot}\!\left(k=\frac{l}{D_A(\chi)},\chi\right) \frac{1}{\chi^2}
\label{eq:ClHT}\ ,\ee
where $P_{\rm h\pdot}\equiv V_{\mu}\left<\delta_{\rm
h}(\bk)\delta_{\pdot}(-\bk)\right>$ is the 3D cross-power spectrum
between halo density fluctuations $\delta_{\rm h}$ and $\pdot$. In
\Eqn{eq:ClHT}, we have included the weight function $\Pi_{ij}$, which
for a mass-selected survey of clusters would have the form:
\be \Pi_{ij}(\chi)\equiv 4\pi D^2_A(\chi)\Theta_{ij}(\chi) \int_{M_{\rm
  min}}^{\infty}dM \frac{n(M,\chi)}{N_{\rm TOT}(\chi_i,\chi_j)} \ , \ee
where $D_{A}$ is the comoving angular diameter distance and
$n(M,\chi)$ the cluster mass function.
%
$N_{\rm TOT}(\chi_i,\chi_j)=\int_{\chi_i}^{\chi_j} d\chi 4\pi
D^2_{A}(\chi) \int_{M_{\rm min}}^{\infty} dM n(M;\chi) \ ,
$
%
is the total number of clusters above mass $M_{\rm min}$. The redshift
shell is selected using the top-hat function:
$\Theta_{ij}(\chi)\equiv\left[\Theta(\chi-\chi_i)-\Theta(\chi-\chi_j)\right]$,
with $\Theta$ the Heaviside step function.
 

Figure~\ref{fig:ClHT} presents the results for $C_l^{\h T}$ between
the CMB and group scale dark matter haloes (left panel) and cluster
mass haloes (right panel). In each case we show the results for 5
narrow bins in redshift, with $\Delta z=0.2$.  We find that on scales
$l<100$ the departures from linear theory are small $<10\%$, and are
characterized by a small amplification of the signal, followed by a
strong suppression. The departures appear small when compared to the
cosmic variance, which is dominated by the $C_l^{TT}$ spectrum.

We next investigated the $\SN$ for $C_l^{\h T}$ and found good
agreement with linear theory expectations: the presence of bias
effectively cancels out in the $\SN$ expression and leads to
negligible changes in the cross-correlation detectability. In fact,
through the increased Poisson noise of the biased sample, there was a
small reduction in the $\SN$, relative to that for the dark
matter--CMB cross-correlation.


\section{Conclusions}

We therefore conclude that the current power spectrum analyses of
\cite{Hoetal2008} and \cite{Giannantonioetal2008} are not affected by
nonlinear density evolution or scale-dependent bias and that these do
not influence the detectability of the ISW-LSS cross-correlation. We
are also led to believe that the results from \cite{Granettetal2008b}
are unlikely to be explained by nonlinear biasing effects in the LCDM
model. Whether there remain systematic errors in the CMB data
associated with point sources or whether this is an exciting detection
of new physics remains an open question.


\vspace{0.2cm}

\noindent {\em Acknowledgments}: RES was supported by a Marie Curie
Reintegration Grant and the Swiss National Foundation under contract
200021-116696/1 and WCU grant R32-2008-000-10130-0.



\input{Blois.RobertSmith.ISW.astro-ph.V1.bbl}
\end{document}

%% file: defs.tex
\newcommand\Eqn[1]     {Eq.\,(\ref{#1})}
\newcommand\Eqns[2]    {Eqs\,(\ref{#1}) and~(\ref{#2})}

\def\mnras{{Mon.~ Not.~ R.~ Astron.~ Soc.~}}

\def\prd{{Phys.~ Rev.~ D.~}}

\def\apj{{Astrophys.~ J.~}}

\def\apjl{{Astrophys.~ J.~ Lett.~}}

\def\nat{{Nature (London)~}}

\def\physrep{{Phys.~ Rep.~}}

\def\be{\begin{equation}}
\def\ee{\end{equation}}
\def\bea{\begin{eqnarray}}
\def\eea{\end{eqnarray}}

\newcommand{\ba}{\begin{eqnarray}}
\newcommand{\ea}{\end{eqnarray}}

\def\nhat{{\hat{\bf n}}}

\def\pp1{{\prime}}
\def\pp2{{\prime\prime}}

\def\2D{{\rm 2D}}

\def\pdot{\dot{\Phi}}

\def\SN{{\mathcal S}/{\mathcal N}}

\def\h{{\rm h}}

\def\bx{{\bf x}}

\def\bk{{\bf k}}
\def\bq{{\bf q}}

\def\1Loop{{\rm 1Loop}}

\def\rhob{\bar{\rho}}

\def\Mpc{\, h^{-1}{\rm Mpc}}

\def\kMpc{\, h \, {\rm Mpc}^{-1}}

\def\dq{d^3\!q}

\def\del{\nabla}

\def\fun#1#2{\lower3.6pt\vbox{\baselineskip0pt\lineskip.9pt
        \ialign{$\mathsurround=0pt#1\hfill##\hfil$\crcr#2\crcr\sim\crcr}}}

%% file: Blois.RobertSmith.ISW.astro-ph.V1.bbl
\begin{thebibliography}{10}

\bibitem{Smithetal2009}
R.~E. {Smith}, C.~{Hern{\'a}ndez-Monteagudo}, and U.~{Seljak}.
\newblock {Impact of scale dependent bias and nonlinear structure growth on the
  integrated Sachs-Wolfe effect: Angular power spectra}.
\newblock {\em \prd}, 80(6):063528--+, September 2009.

\bibitem{SachsWolfe1967}
R.~K. {Sachs} and A.~M. {Wolfe}.
\newblock {Perturbations of a Cosmological Model and Angular Variations of the
  Microwave Background}.
\newblock {\em \apj}, 147:73--+, January 1967.

\bibitem{ReesSciama1968}
M.~J. {Rees} and D.~W. {Sciama}.
\newblock {Larger scale Density Inhomogeneities in the Universe}.
\newblock {\em \nat}, 217:511--+, February 1968.

\bibitem{BirkinshawGull1983}
M.~{Birkinshaw} and S.~F. {Gull}.
\newblock {A test for transverse motions of clusters of galaxies}.
\newblock {\em \nat}, 302:315--317, March 1983.

\bibitem{CrittendenTurok1996}
R.~G. {Crittenden} and N.~{Turok}.
\newblock {Looking for a Cosmological Constant with the Rees-Sciama Effect}.
\newblock {\em Physical Review Letters}, 76:575--578, January 1996.

\bibitem{Giannantonioetal2008}
T.~{Giannantonio}, R.~{Scranton}, R.~G. {Crittenden}, R.~C. {Nichol}, S.~P.
  {Boughn}, A.~D. {Myers}, and G.~T. {Richards}.
\newblock {Combined analysis of the integrated Sachs-Wolfe effect and
  cosmological implications}.
\newblock {\em \prd}, 77(12):123520--+, June 2008.

\bibitem{Hoetal2008}
S.~{Ho}, C.~{Hirata}, N.~{Padmanabhan}, U.~{Seljak}, and N.~{Bahcall}.
\newblock {Correlation of CMB with large-scale structure. I. Integrated
  Sachs-Wolfe tomography and cosmological implications}.
\newblock {\em \prd}, 78(4):043519--+, August 2008.

\bibitem{Granettetal2008b}
B.~R. {Granett}, M.~C. {Neyrinck}, and I.~{Szapudi}.
\newblock {An Imprint of Superstructures on the Microwave Background due to the
  Integrated Sachs-Wolfe Effect}.
\newblock {\em \apjl}, 683:L99--L102, August 2008.

\bibitem{Bernardeauetal2002}
F.~{Bernardeau}, S.~{Colombi}, E.~{Gazta{\~n}aga}, and R.~{Scoccimarro}.
\newblock {Large-scale structure of the Universe and cosmological perturbation
  theory}.
\newblock {\em \physrep}, 367:1--3, September 2002.

\bibitem{Springel2005}
V.~{Springel}.
\newblock {The cosmological simulation code GADGET-2}.
\newblock {\em \mnras}, 364:1105--1134, December 2005.

\bibitem{Smith2009}
R.~E. {Smith}.
\newblock {Covariance of cross-correlations: towards efficient measures for
  large-scale structure}.
\newblock {\em \mnras}, pages 1337--+, September 2009.

\end{thebibliography}
